\documentclass[aps,prl,twocolumn,showpacs]{revtex4}
\usepackage{hyperref}

\newcommand{\Label}[1]{\label{#1}}
\newcommand{\GPE}{Gross-Pitaevskii equation}

\newcommand{\etal}{{\em et al.}}
\newcommand{\pt}[1]{\par\noindent{\bf #1 }}

\def\DRAFT{%
\renewcommand{\Label}[1]{\label{##1}
{\hbox to 0cm{\textcolor{magenta}{\hss\em ##1\quad}}}}}
\usepackage{graphicx}
\usepackage{dcolumn}
\usepackage{amsmath}
\usepackage{color}
\usepackage{times}
\newcommand{\btimes}{\boldsymbol{\times}}
\newcommand{\bcdot}{\boldsymbol{\cdot}}
\begin{document}
\title{The nucleation, growth and stabilization of vortex lattices}
\author{A. A. Penckwitt}
\affiliation{Department of Physics, University of Otago, Dunedin, New
Zealand} 
\author{R. J. Ballagh}
\affiliation{Department of Physics, University of Otago, Dunedin, New
Zealand}
\author{C. W. Gardiner}
\affiliation{School of Chemical and Physical Sciences, Victoria University,
Wellington, New Zealand}
\begin{abstract} 
We give a simple unified theory of vortex nucleation and vortex lattice 
formation which is valid from the initiation process up to the final 
stabilization of the lattice.  We treat the growth of vortex lattices from a
rotating thermal cloud, and their production using a rotating trap. We find
results consistent with previous work on the critical velocity or critical
angular velocity for vortex formation, and predict the initial number of
vortices expected before their self assembly into a lattice. We show that the
thermal cloud plays a crucial role in the process of vortex  lattice
nucleation.
\end{abstract}

\pacs{03.75.Fi} 
\maketitle

The opportunity that Bose Einstein condensates present for precisely controlled
experimental and theoretical study of superfluids has revived an  intense
interest in the properties and behavior of quantized vortices. Many different
theoretical calculations have been made of stationary single vortex states and
their excitations (\cite{Fetter2001c} and references therein), and vortex
lattices  (\cite{Feder1999b} and references therein) in Bose condensates, but
the recent experimental observations of vortex lattices
\cite{Madison2000a,Madison2001a,Haljan2001a,Raman2001ax,Abo-Shaeer2001b,%
Abo-Shaeer2001a,Hodby2001a} have  focussed attention on the {\em mechanisms}
of vortex formation. 

One experimental method for creating vortices is to stir a condensate with an
anisotropic potential,  and a number of theoretical analyses of this scenario
have been made in terms of  the \GPE, e.g.\  
\cite{Fetter2001a,Dalfovo2000a,Anglin2001a,Muryshev2001a}. The view of these
treatments is that  the perturber causes a mixing of  the condensate ground
state and excited condensate states (with angular  momentum values determined
by the stirrer geometry). Typically, a perturbative  calculation is used to
obtain a critical rotational speed (or a critical linear speed at the
Thomas-Fermi radius) of the stirrer at which the mixing becomes effective. The
argument is then made that an instability will lead to growth of the vortex
state.  A more complete non-perturbative calculation for a  localized rotating
stirrer \cite{Caradoc-Davies1999a} shows that  in this case  the mixing is
predominantly between the ground state and an $l=1$  vortex state, and that as
the relative amplitudes evolve by coherent ``Rabi cycling'', the full
condensate exhibits a  vortex cycling from infinity to the central regions of
the condensate.  No coherent mixing mechanisms, however, can explain the
formation of a {\em vortex lattice}, since an energy barrier exists between the
superposition  state and the vortex lattice state, and some additional
mechanism is required to remove the energy liberated when the vortex lattice is
formed. 

An alternative method of producing vortices, in which a vapor of cold atoms is 
evaporatively cooled so as to preserve its angular momentum  has been 
demonstrated by Haljan  \etal\ \cite{Haljan2001a}. In this experiment, which
involves no stirring  potential, the mechanism of formation of  the vortices
must involve a modification of the theory of condensate  growth (which now
exists in a  reasonably good quantitative form; \cite{Davis2001a} and
references therein) to take account of  the non-zero angular
momentum of the vapor from which the condensate is formed. In this paper we
will show that the mechanism thus demonstrated is the fundamental
process in those experiments in which the stirring of the condensate apparently
generates a vortex lattice. Weak stirring, in the absence of any thermal cloud,
only produces ``Rabi cycling'' of angular momentum in and out of the
condensate. However, if the stirring also produces a rotating thermal cloud,
then, by the same process responsible for the growth of a condensate
\cite{Gardiner1997b}, angular momentum is transferred {\em irreversibly} into 
the condensate, leading in equilibrium to a rotating vortex lattice. In 
general terms, this has already been noted in \cite{Zhuravlev2001a}, and in 
\cite{Gardiner2002a} we derived a very simplified vortex growth 
equation---given below as Eq.~(\ref{1600})---embodying the essentials of this 
concept.

Other theoretical treatments of condensate dynamics at finite temperature have 
been applied to the case of vortex dynamics.  Zhuravlev  \etal\
\cite{Zhuravlev2001a} studied the interaction of vortex lattices  with a thermal
cloud in the presence of a trap asymmetry, and from   broad conservation
principles they  obtained equations for the time evolution  of the angular
velocity of the condensate and thermal cloud, and the transfer  of energy
between them. Williams \etal\ \cite{Williams2002a} showed how  the inclusion of
interactions with a thermal vapor cloud would give rise to an  instability of 
surface modes, as required to produce vortices.  Our own treatment
\cite{Gardiner2002a} had a basis  broadly equivalent to that of Williams \etal,
but showed how  the interaction with the vapor cloud  could produce an equation
which should describe the full process of vortex  nucleation, growth and
stabilization.  Using this equation, we will show in  this paper how the growth
of vortex lattices can be understood as a condensate  growth process, and that
the condition that the gain in this growth process be  positive is the same as
the criteria noted above for vortex  nucleation. We also 
demonstrate explicitly how the vortices  produced in the ``Rabi cycling''
caused by a rotating trap nucleate the growth  process, which then proceeds to
dominate the vortex lattice formation process.

\pt{The \GPE} for
$ \psi_R = \exp(i\,\boldsymbol{\Omega}\bcdot{\bf L}t/\hbar)\psi $, the
condensate wavefunction in a frame of reference rotating with angular velocity 
 $ \boldsymbol{\Omega}$,  takes the form
\begin{eqnarray}\Label{vl1}
i\hbar\frac{\partial\psi_R}{\partial t} = 
\left(-\frac{\hbar^2}{2m}\nabla^2
+V_T^R
+ u \bigl |\psi_R\bigr |^2 
 -\boldsymbol{\Omega}\bcdot{\bf L}\right)\psi_R,
\end{eqnarray}
where $V_T^R({\bf x})$ is the trap potential in the rotating frame. A vortex
lattice results from the  stationary solutions of Eq.~(\ref{vl1}) satisfying $
i\hbar{\partial\psi_R/\partial t} =\mu_C \psi_R$  \cite{Feder1999a,Feder1999b},
which are normally obtained by integrating the \GPE\ along an imaginary time
direction \cite{Feder1999a}. Indeed, noting that  the achievement of the 
vortex lattice stationary state requires dissipation in order to allow the 
system to settle into the energy minimum associated  with the lattice, Tsubota
\etal\ \cite{Tsubota2002a}  proposed an {\em ad hoc}  modification of this
method as a phenomenological model of the process of  vortex lattice formation. 
They modified the left-hand side of Eq.~(\ref{vl1}) to  $(1+i\gamma)
i\hbar{\partial\psi_R/\partial t}$, and compensated for the consequent loss 
of probability by continuously adjusting the norm of the wavefunction. 

\pt{Origin of dissipation:} When the condensate is grown from a rotating  vapor
cloud the growth process itself is dissipative, and in the  experiments of
\cite{Madison2000a,Madison2001a,Haljan2001a,Raman2001ax} dissipation can arise 
by transfer of atoms between the thermal cloud and the condensate. The vortex 
growth equation derived in \cite{Gardiner2002a} uses
quantum kinetic theory for the growth of  the wavefunction for a  condensate
trapped in a potential $ V_T({\bf x},t)$ rotating with angular  velocity $
\boldsymbol{\Omega}$,  from a thermal cloud  held at temperature $ T$ and
chemical potential  $ \mu$ and rotating with angular velocity
$\boldsymbol{\alpha}$. A dissipative term  arises from  collisions between
atoms in a thermal cloud of atoms trapped by the same  potential as the
condensate, in which one of the colliding atoms enters the  condensate after
the collision.  The net rate at which atoms enter and leave  the  condensate as
the result of such collisions  is  characterized by a coefficient 
\begin{subequations}
\begin{eqnarray}\Label{3}
W^+({\bf x}) &=& \frac{u^2}{(2\pi)^5\hbar^2}
\int d^3{\bf K}_1 d^3{\bf K}_2 d^3{\bf K}_3 \,
\delta\left(\omega_{123}\right)
\nonumber\\&&\times
\delta({\bf K}_1+{\bf K}_2 -{\bf K}_3 )
F_1F_2(1+F_3)
\\ \Label{3a}
&\approx& g\times{4m(akT)^2/\pi\hbar^3},
\end{eqnarray}
\end{subequations}
and in practice $ g \approx 3$ fits most growth experiments. Here $ F({\bf
x},{\bf K})$ is the Bose-Einstein distribution function for  a thermal cloud
in  an effective potential in the frame rotating at the  {\em cloud's} angular
velocity $\boldsymbol{\alpha}$, i.e.\ $ V_{\alpha}({\bf x}) \equiv\bar
V_T({\bf x}) - {m\left(\boldsymbol{\alpha}\btimes{\bf x} \right)^2/2}$, where $
\bar V_T({\bf x})$ is the time averaged trap potential  over the differential
rotation between the trap and the  cloud. The vortex growth equation takes the
form in the frame rotating with the  {\em trap potential}
\begin{eqnarray}\Label{1600}
i\hbar\frac{\partial\psi_R}{\partial t}
&=&\left(-\frac{\hbar^2}{2m}\nabla^2
+V_T^R({\bf x}) + u \bigl |\psi_R\bigr |^2
 -\boldsymbol{\Omega}\bcdot{\bf L}\right)\psi_R
\nonumber \\ 
&&\quad+ i\gamma
\left(\mu+(\boldsymbol{\alpha}-\boldsymbol{\Omega})\bcdot{\bf L}
- i\hbar\frac{\partial}{\partial t}\right)\psi_R,
\end{eqnarray}
where $\gamma\equiv \hbar W^{+}/kT$.  It is important to distinguish between
the {\em cloud's} angular velocity $  \boldsymbol{\alpha}$, and the {\em
trap's} angular velocity  $ \boldsymbol{\Omega}$.  If these are different the
distribution of the thermal  cloud is not truly thermodynamic equilibrium, and
the form chosen represents  an  approximation expressing the experimental
observation that it is very difficult to spin up a thermal  cloud by means of a
rotating potential.  One therefore expects that    $\alpha <\Omega$. In this
paper, however, we shall consider mainly the cases in which either the trap is
cylindrically symmetric, so that we may set $  \boldsymbol{\Omega}=0$, or in
which the cloud and the trap rotate at the same velocity, so $
\boldsymbol{\alpha}=\boldsymbol{\Omega}$. Eq.~(\ref{1600})  is essentially
equivalent in terms of the physical  assumptions required for its derivation to
that of \cite{Williams2002a},  although  its appearance is very different, and
it is in practice easier to solve. In  contrast, although Eq.~(\ref{1600})
looks superficially very similar to that of  \cite{Tsubota2002a}, it is in fact
very different, as explained in  \cite{Gardiner2002a}.

\begin{figure}[t]
\includegraphics[width=6.7cm]{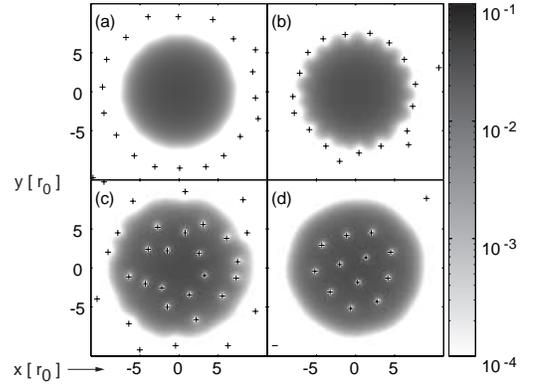}
\vskip -4mm
\caption{Condensate density during formation of vortex lattice 
from a
rotating thermal cloud. Vortices are marked by $+$ or $-$ according
to their sense. (a) $t=29.28\,t_0$, (b) $t=37.70\,t_0$, (c) $t=52.40\,t_0$, 
(d) $t=187.24\,t_0$. 
Condensate wavefunction seeded as described in text. Parameters are 
$\alpha = 0.65\,\omega$, $\mu = 12\hbar\omega $, $\gamma = 0.1$, $u = 1000 u_0$, 
initial $\mu_C = 12.7 \hbar\omega $. Units are: time $t_0 = 1/\omega$, distance
 $r_0 = \sqrt{\hbar/2m\omega}$, 
 collisional strength $u_0 = \hbar \omega r_0^2$.}
\label{seeddens.eps}
\vskip -6mm
\end{figure}
\pt{Simulation results for a rotating vapor cloud:}
We have simulated Eq.~(\ref{1600}) in two dimensions for many different
scenarios, and in Fig.~\ref{seeddens.eps} we present a time sequence of
results from a representative case in which the condensate is initially in the
ground state of a rotationally symmetric harmonic trap of frequency  $\omega
$,  the thermal cloud has $\mu =12\hbar \omega $, and  $\alpha=0.65\,\omega $.
The initial condensate has unit norm and chemical potential
$\mu_C=12.7\hbar \omega$, although they can be any non-zero number.  We
simulate the non-stimulated collisions which start the process by adding to the
intial wavefunction a uniform superposition of angular momentum states with
$l=1$ to $30$, on a Gaussian radial profile centered at the Thomas-Fermi radius
and with maximum amplitude $\sim 2 \times 10^{-7}$. 

Fig.~\ref {seeddens.eps}(a) is a density plot of the condensate near the end
of the first stage of the evolution, in which an imperfect ring of $19$
vortices arrives from infinity to just outside the Thomas-Fermi radius. As this
ring shrinks further (Fig.~\ref{seeddens.eps}(b)), several vortices are shed,
and a dominant ring of 16 vortices passes through the Thomas-Fermi
radius into the interior of the condensate. These vortices then distribute
themselves irregularly but quasi-uniformly over the condensate which expands
and picks up angular velocity (Fig.~\ref{seeddens.eps}(c)). Subsequently, over
a long period, further vortices leave (and for larger values of $\alpha$ may
enter) the dense region, until finally a regular lattice of 12 vortices rotating
at angular velocity $ \alpha $ in the lab frame remains 
(Fig.~\ref{seeddens.eps}(d)). 

\begin{figure}[t]
\includegraphics[height=3cm]{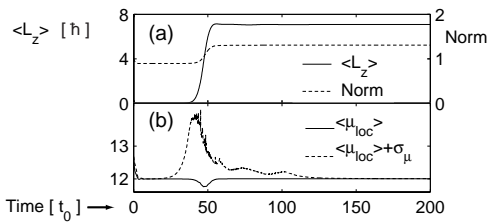}
\vskip -4mm
\caption{
Time evolution of condensate quantities for the case of 
Fig.~\ref{seeddens.eps}. (a) $\langle L_z\rangle$ and norm, 
(b) $ \langle\mu_{\mathrm{loc}}\rangle$ and $ \sigma_{\mu}$ (in units of $\hbar \omega$).}
\label{seedLmuevol.eps}
\vskip -6mm
\end{figure}

The overall time scale of the process is illustrated in 
Fig.~\ref{seedLmuevol.eps}(a) which shows the total angular momentum increasing
sharply around $t\approx 40$, when the 16 vortices enter the dense part of the
condensate. For a given seeding, the  timescale for this onset is determined by
$\gamma$ and to good approximation scales as $\gamma ^{-1}.$ The  {\em local
chemical potential $\mu _{\mathrm{loc}}({\bf x})$} (defined as the absolute 
value of the right-hand side of
Eq.~(\ref{vl1}) divided by $\psi _{R}$) provides a useful characterisation of 
the temporal development. In Fig.~{\ref{seedLmuevol.eps}(b) }, we plot the
evolution of the spatial mean of $\mu _{\mathrm{loc}}({\bf x})$, which 
exhibits a rapid initial adjustment (to $\approx \mu$), as the condensate 
roughly equilibrates with the thermal cloud, and a later dip at $ t\approx 48$ 
during which the number of atoms in the condensate increases (see 
Fig.~\ref{seedLmuevol.eps}(a)). The spatial variance $\sigma_{\mathrm{\mu}}$ 
of $ \mu _{\mathrm{loc}}({\bf x})$
quantifies the deviation of the solution from equilibrium (where it is zero).
It becomes large as the vortices enter the dense region of the condensate, and
its subsequent slow decay indicates the long time needed for stabilization of
the vortex lattice. It is not easy to characterize the latter time scale, since
the local chemical potential can be almost constant everywhere, with possibly
only a single vortex not in its final place. From a large number of simulations
we have found that the number of vortices in the stable final lattice increases
with $\alpha $ and $\mu $, and is generally different from the number seen in
the initial ring. It also exhibits a very weak dependence on the exact form of
the seeding. For the case of $\mu =12\hbar \omega $ the critical value for the
appearance of any vortices is found from the simulations to be $\alpha
= 0.444\,\omega $, at which speed an irregular ring of $10$ vortices appears 
initially, settling down eventually to three vortices in a lattice inside the
Thomas-Fermi radius. Three dimensional simulations of Eq.~(\ref{1600}) also 
confirm the general behavior seen in Fig.~\ref{seeddens.eps}.

\pt{Analytic treatment:} An analytic understanding of
the nucleation and growth process of the vortices can be obtained from 
Eq.~(\ref{1600}) by considering the linearized Bogolyubov excitations above the
initial state, which have angular momentum $l$ and energy eigenvalues
$\epsilon _{n,l}$ measured relative to $\mu_C$.  From the final term of 
Eq.~(\ref{1600}) we see that excitations for which $\epsilon _{n,l}<\mu -\mu
_{C}+\hbar\alpha l$ will experience a positive growth rate $G\approx \gamma
(\mu -\mu _{C}+\hbar\alpha l-\epsilon _{n,l})/\hbar.$ Initially this causes 
rapid growth of the condensate (the $n=l=0$ component, which has the lowest
eigenvalue) until the condensate and thermal cloud approximately equilibrate
$\left( \mu \approx \mu _{C}\right) $. Subsequently, the gain of the other
components is $\gamma (\alpha l - \epsilon _{n,l}/\hbar)$, and thus the critical
value of $\alpha $ for positive gain is given by $\epsilon _{0,l}=\hbar\alpha
l,$ the same condition  for criticality found in \cite{Williams2002a}. Above
criticality, the dominant value $l_{v}$ is that which maximizes the gain, and
is given by $\partial (\hbar\alpha l - \epsilon _{0,l})/\partial l=0$, and a
crude estimate for $l_{v}$ can be obtained by minimizing
$\hbar^2l^{2}/2mr^{2}-\hbar\alpha l$ at the  Thomas-Fermi surface, to give
$l_{v}\approx {\pi R_{{\rm TF}}^{2}2m\alpha /h}$. This is the number of
vortices which would result from filling a disk of radius $R_{ {\rm TF}}$ with
a vortex lattice \cite{Donnelly1991a}. In the initial stages of vortex
formation, when the process is still linear, the wavefunction of the condensate
takes the form ( for simplicity including only the excitation with maximum
gain, which has angular momentum $\hbar l_{v}$ and energy $ \hbar \omega
_{v}\equiv \hbar\alpha l_v - \epsilon_{0,l_v}$) 
\begin{eqnarray}
\Label{vl17}
\psi_R \approx e^{-i\mu _{C}t/\hbar }\bigg\{\xi _{0}+e^{il_{v}\phi
+Gt}\left[ ue^{i\omega _{v}t}+ve^{-i\omega _{v}t}\right] \bigg\}\;. 
\end{eqnarray}
Here $\xi _{0}(r)$ is the initial rotationally symmetric condensate
wavefunction, $\phi $ is the azimuthal angle, and $u(r),v(r)$ are obtained by
solving the Bogolyubov-de Gennes equations. The essential behavior can be seen
by neglecting $v(r)$, so that $\psi $ has $l_{v}$ zeroes---that is
vortices---given by the interference of the two terms. These vortices all occur
at the same radius, initially at infinity. Since the long distance behavior of
$u(r)$ must be less rapid than that of $\xi _{0}(r)$, the ring will steadily
shrink as the excitation grows. The time dependence of the coefficient means
that the ring of vortices rotates at the angular frequency  $\omega _{v}$ in
the rotating frame. In a more detailed picture, more values of $l$ should be
included, and the corresponding superposition will be an imperfectly circular
ring of vortices, perhaps containing more than $l_{v}$ vortices.

\begin{figure}[t]
\includegraphics[height=3.5cm]{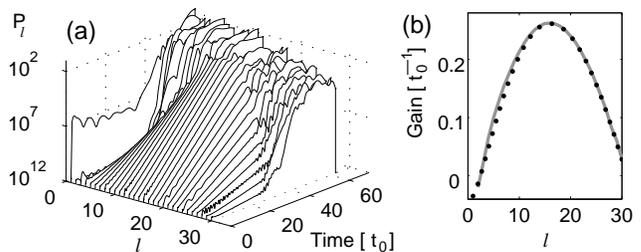}
\vskip -5mm
\caption{\label{meshproj0.1.eps}
(a) Evolution of angular momentum occupation probabilities $P_l$ for $l=1$ to 
$30$ for the case of Fig.~\ref{seeddens.eps}. (b) Comparison of gain 
coefficients of $l$ components obtained from the simulation in 
Fig.~\ref{seeddens.eps} (solid line) with the predicted gain coefficients
$G=\gamma (\alpha l - \epsilon_{0,l}/\hbar)$ (dotted line).}
\vskip -5mm
\end{figure}

This analytic treatment can be verified by decomposing the condensate into
angular momentum components, and plotting the occupation values $P_{l}$
against time, as in Fig.~\ref{meshproj0.1.eps}(a). 
We see that the $l$ components above threshold grow exponentially up to the
time $t\approx 40$ (when the vortex ring passes through the Thomas-Fermi
radius), and the gain is maximum for $l_{v}=16$. In  
Fig.~\ref{meshproj0.1.eps}(b) we compare the gain prediction $\gamma (\alpha l 
- \epsilon_{0,l}/\hbar)$ (obtained by calculating the surface modes $\epsilon
_{0,l}$ for a condensate with $\mu _{C}=12\hbar \omega $ ) against the gain
rates measured from the simulation. Not only is our gain prediction very
accurate, but we also find that the spatial particle density for each $l$
component very accurately matches the prediction from the corresponding
Bogolyubov wavefunction.

\begin{figure}[t]
\includegraphics{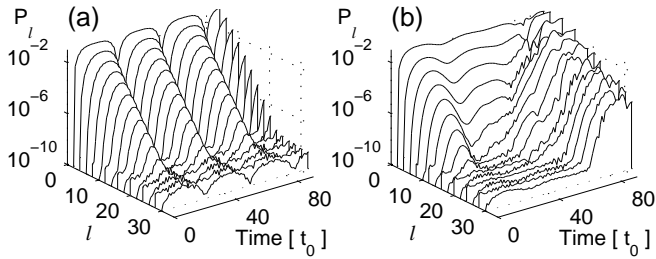}
\vskip -4mm
\caption{\label{fig-rot.eps}
Evolution of angular momentum occupation probabilities $P_l$ for: (a) rotating 
elliptical trap with no thermal cloud ($\gamma = 0$); (b) rotating 
elliptical trap and corotating thermal cloud. Only the even $l$ states are 
shown. Parameters as in Fig.~\ref{seeddens.eps} except trap parameters 
$\omega_x=1.05\,\omega$, $\omega_y=1.15\,\omega$ and 
$\alpha = \Omega = 0.65\,\omega$.}
\vskip -6mm
\end{figure}

\pt{Rotating trap, no thermal cloud:} Finally, we consider the connection
between stirring and growth. As an example of a case of pure stirring we have
simulated a rotating elliptical trap in the absence of a thermal cloud. The
general behavior is well represented by the case $\mu _{C}=12.7\hbar \omega $
and $\Omega =0.65\,\omega$, for which we find that the condensate deforms into
an elliptical shape rotating at the trap frequency, and a similarly deformed
ring of about 28 vortices shrinks until it meets the dense part of the
condensate at its narrowest radius. The resulting state is a distorted version
of Fig.~\ref{seeddens.eps}(b) \cite{Penckwitt2002a}, and this subsequently
undergoes an oscillatory behavior, but with no further penetration of the
vortices into the condensate. Only the even angular momentum components of
the condensate become significantly occupied and undergo periodic cycling 
(Fig.~\ref{fig-rot.eps}(a)). This
can be well understood in terms of the Rabi cycling model 
\cite{Caradoc-Davies1999a}, noting that an elliptical potential can connect the
initial ground state only to states of even $l$. Resonant mixing to the surface
modes will occur when $\hbar\Omega l - \epsilon_{0,l} = 0$, and the lowest
value of $ \Omega$ for which this occurs  (when also 
$\partial(\hbar\Omega l-\epsilon_{0,l})/\partial l = 0 $) gives the same
condition for criticality as found in
Refs.~\cite{Dalfovo2000a,Anglin2001a,Muryshev2001a}. This is also exactly the
same condition as for the rotating cloud.
\pt{Rotating trap with a thermal cloud:} Now repeating this simulation, but 
with a thermal cloud (Eq.~(\ref{1600}) with $
\alpha=\Omega$ and $ \gamma \ne 0$), we get dramatically different results.  The
same elliptical distortion of the condensate occurs, but the effect tends to
smooth out with time. A distorted vortex ring appears, with an initial number
of vortices as expected from the growth  mechanism (here $\approx 14$), and then
shrinks into the condensate and nucleates the vortex lattice. The angular
momentum projections in Fig.~\ref{fig-rot.eps}(b) show the difference very
clearly.  The vortex lattice is  the result of the growth process, which is
essentially superposed on the  effects of stirring, eventually masking them.
This scenario leads to the conclusion that the principal effect of  the
rotating  trap is to produce a rotating thermal cloud, which then
transfers angular  momentum into the condensate, eventually forming the
vortex lattice.

\pt{Comparison with other work:} The main point of this paper is the analysis
of vortex formation in terms of the  growth in occupation of modes with non-zero
angular momentum. This requires the presence of a rotating thermal cloud, but
stirring the condensate can seed angular momentum components which may then
grow from the thermal cloud by stimulated collisions. The critical angular
velocity of the thermal cloud to allow growth is the same as the critical
angular velocity required for stirring to resonantly mix surface modes with the
condensate, and agrees with the latter values calculated by previous workers
\cite{Dalfovo2000a,Anglin2001a,Muryshev2001a,Williams2002a}. However, our
treatment goes  beyond the critical point, after which the behavior of the two
processes becomes quite different, and includes the full dynamics of lattice
stabilization.  Our vortex growth equation (\ref{1600}) can  be solved without
linearization or using hydrodynamic methods, and its  solutions  can be
interpreted in a straightforward quantum mechanical way.  The dynamics  of the
thermal cloud are suppressed in this paper,  but their inclusion is 
necessary for the full understanding of  vortex  lattice dynamics, including
lattice decay, which can also be treated by our formalism. 

\begin{acknowledgments}
This work was supported by the Marsden Fund of New Zealand under contract
PVT-902.
\end{acknowledgments}

\vskip -2mm

\bibliography{VortexLetter}

\begin{thebibliography}{23}
\expandafter\ifx\csname natexlab\endcsname\relax\def\natexlab#1{#1}\fi
\expandafter\ifx\csname bibnamefont\endcsname\relax
  \def\bibnamefont#1{#1}\fi
\expandafter\ifx\csname bibfnamefont\endcsname\relax
  \def\bibfnamefont#1{#1}\fi
\expandafter\ifx\csname citenamefont\endcsname\relax
  \def\citenamefont#1{#1}\fi
\expandafter\ifx\csname url\endcsname\relax
  \def\url#1{\texttt{#1}}\fi
\expandafter\ifx\csname urlprefix\endcsname\relax\def\urlprefix{URL }\fi
\providecommand{\bibinfo}[2]{#2}
\providecommand{\eprint}[2][]{\url{#2}}

\bibitem[{\citenamefont{Fetter and Svidinsky}(2001)}]{Fetter2001c}
\bibinfo{author}{\bibfnamefont{A.~L.} \bibnamefont{Fetter}} \bibnamefont{and}
  \bibinfo{author}{\bibfnamefont{A.~A.} \bibnamefont{Svidinsky}},
  \bibinfo{journal}{J. Phys. Condens. Matter} \textbf{\bibinfo{volume}{13}},
  \bibinfo{pages}{R135} (\bibinfo{year}{2001}).

\bibitem[{\citenamefont{Feder et~al.}(1999{\natexlab{a}})\citenamefont{Feder,
  Clark, and Schneider}}]{Feder1999b}
\bibinfo{author}{\bibfnamefont{D.~L.} \bibnamefont{Feder}},
  \bibinfo{author}{\bibfnamefont{C.~W.} \bibnamefont{Clark}}, \bibnamefont{and}
  \bibinfo{author}{\bibfnamefont{B.~I.} \bibnamefont{Schneider}},
  \bibinfo{journal}{Phys. Rev. A} \textbf{\bibinfo{volume}{61}},
  \bibinfo{pages}{011601(R)} (\bibinfo{year}{1999}{\natexlab{a}}).

\bibitem[{\citenamefont{Madison et~al.}(2000)\citenamefont{Madison, Chevy,
  Wohlleben, and Dalibard}}]{Madison2000a}
\bibinfo{author}{\bibfnamefont{K.~W.} \bibnamefont{Madison}},
  \bibinfo{author}{\bibfnamefont{F.}~\bibnamefont{Chevy}},
  \bibinfo{author}{\bibfnamefont{W.}~\bibnamefont{Wohlleben}},
  \bibnamefont{and} \bibinfo{author}{\bibfnamefont{J.}~\bibnamefont{Dalibard}},
  \bibinfo{journal}{Phys. Rev. Lett.} \textbf{\bibinfo{volume}{84}},
  \bibinfo{pages}{806} (\bibinfo{year}{2000}).

\bibitem[{\citenamefont{Madison et~al.}(2001)\citenamefont{Madison, Chevy, and
  Dalibard}}]{Madison2001a}
\bibinfo{author}{\bibfnamefont{K.~W.} \bibnamefont{Madison}},
  \bibinfo{author}{\bibfnamefont{F.}~\bibnamefont{Chevy}}, \bibnamefont{and}
  \bibinfo{author}{\bibfnamefont{J.}~\bibnamefont{Dalibard}},
  \bibinfo{journal}{Phys. Rev. Lett.} \textbf{\bibinfo{volume}{86}},
  \bibinfo{pages}{4443} (\bibinfo{year}{2001}).

\bibitem[{\citenamefont{Haljan et~al.}(2001)\citenamefont{Haljan, Coddington,
  Engels, and Cornell}}]{Haljan2001a}
\bibinfo{author}{\bibfnamefont{P.~C.} \bibnamefont{Haljan}},
  \bibinfo{author}{\bibfnamefont{I.}~\bibnamefont{Coddington}},
  \bibinfo{author}{\bibfnamefont{P.}~\bibnamefont{Engels}}, \bibnamefont{and}
  \bibinfo{author}{\bibfnamefont{E.~A.} \bibnamefont{Cornell}},
  \bibinfo{journal}{Phys. Rev. Lett.} \textbf{\bibinfo{volume}{87}},
  \bibinfo{pages}{210403} (\bibinfo{year}{2001}).

\bibitem[{\citenamefont{Raman et~al.}(2001)\citenamefont{Raman, Abo-Shaeer,
  Vogels, Xu, and Ketterle}}]{Raman2001ax}
\bibinfo{author}{\bibfnamefont{C.}~\bibnamefont{Raman}},
  \bibinfo{author}{\bibfnamefont{J.~R.} \bibnamefont{Abo-Shaeer}},
  \bibinfo{author}{\bibfnamefont{J.~M.} \bibnamefont{Vogels}},
  \bibinfo{author}{\bibfnamefont{K.}~\bibnamefont{Xu}}, \bibnamefont{and}
  \bibinfo{author}{\bibfnamefont{W.}~\bibnamefont{Ketterle}},
  \bibinfo{journal}{Phys. Rev. Lett.} \textbf{\bibinfo{volume}{87}},
  \bibinfo{pages}{210402} (\bibinfo{year}{2001}).

\bibitem[{\citenamefont{Abo-Shaeer et~al.}(2001)\citenamefont{Abo-Shaeer,
  Raman, Vogels, and Ketterle}}]{Abo-Shaeer2001b}
\bibinfo{author}{\bibfnamefont{J.~R.} \bibnamefont{Abo-Shaeer}},
  \bibinfo{author}{\bibfnamefont{C.}~\bibnamefont{Raman}},
  \bibinfo{author}{\bibfnamefont{J.~M.} \bibnamefont{Vogels}},
  \bibnamefont{and} \bibinfo{author}{\bibfnamefont{W.}~\bibnamefont{Ketterle}},
  \bibinfo{journal}{Science} \textbf{\bibinfo{volume}{292}},
  \bibinfo{pages}{476} (\bibinfo{year}{2001}).

\bibitem[{\citenamefont{Abo-Shaeer et~al.}(2002)\citenamefont{Abo-Shaeer,
  Raman, and Ketterle}}]{Abo-Shaeer2001a}
\bibinfo{author}{\bibfnamefont{J.~R.} \bibnamefont{Abo-Shaeer}},
  \bibinfo{author}{\bibfnamefont{C.}~\bibnamefont{Raman}}, \bibnamefont{and}
  \bibinfo{author}{\bibfnamefont{W.}~\bibnamefont{Ketterle}},
  \bibinfo{journal}{Phys. Rev. Lett.} \textbf{\bibinfo{volume}{88}},
  \bibinfo{pages}{070409} (\bibinfo{year}{2002}).

\bibitem[{\citenamefont{Hodby et~al.}(2001)\citenamefont{Hodby, Heckenblaikner,
  Hopkins, Marag{\'o}, and Foot}}]{Hodby2001a}
\bibinfo{author}{\bibfnamefont{E.}~\bibnamefont{Hodby}},
  \bibinfo{author}{\bibfnamefont{G.}~\bibnamefont{Heckenblaikner}},
  \bibinfo{author}{\bibfnamefont{S.~A.} \bibnamefont{Hopkins}},
  \bibinfo{author}{\bibfnamefont{O.~M.} \bibnamefont{Marag{\'o}}},
  \bibnamefont{and} \bibinfo{author}{\bibfnamefont{C.}~\bibnamefont{Foot}},
  \bibinfo{journal}{Phys. Rev. Lett.} \textbf{\bibinfo{volume}{88}},
  \bibinfo{pages}{010405} (\bibinfo{year}{2001}).

\bibitem[{\citenamefont{Fetter}(2001)}]{Fetter2001a}
\bibinfo{author}{\bibfnamefont{A.~L.} \bibnamefont{Fetter}},
  \bibinfo{journal}{Phys. Rev. A} \textbf{\bibinfo{volume}{64}},
  \bibinfo{pages}{063608} (\bibinfo{year}{2001}).

\bibitem[{\citenamefont{Dalfovo and Stringari}(2000)}]{Dalfovo2000a}
\bibinfo{author}{\bibfnamefont{F.}~\bibnamefont{Dalfovo}} \bibnamefont{and}
  \bibinfo{author}{\bibfnamefont{S.}~\bibnamefont{Stringari}},
  \bibinfo{journal}{Phys. Rev. A.} \textbf{\bibinfo{volume}{63}},
  \bibinfo{pages}{011601} (\bibinfo{year}{2000}).

\bibitem[{\citenamefont{Anglin}(2001)}]{Anglin2001a}
\bibinfo{author}{\bibfnamefont{J.~R.} \bibnamefont{Anglin}},
  \bibinfo{journal}{Phys. Rev. Lett.} \textbf{\bibinfo{volume}{87}},
  \bibinfo{pages}{240401} (\bibinfo{year}{2001}).

\bibitem[{\citenamefont{Muryshev and Fedichev}()}]{Muryshev2001a}
\bibinfo{author}{\bibfnamefont{A.~E.} \bibnamefont{Muryshev}} \bibnamefont{and}
  \bibinfo{author}{\bibfnamefont{P.~O.} \bibnamefont{Fedichev}},
  \eprint{cond-mat/0106462}.

\bibitem[{\citenamefont{Caradoc-Davies
  et~al.}(1999)\citenamefont{Caradoc-Davies, Ballagh, and
  Burnett}}]{Caradoc-Davies1999a}
\bibinfo{author}{\bibfnamefont{B.~M.} \bibnamefont{Caradoc-Davies}},
  \bibinfo{author}{\bibfnamefont{R.~J.} \bibnamefont{Ballagh}},
  \bibnamefont{and} \bibinfo{author}{\bibfnamefont{K.}~\bibnamefont{Burnett}},
  \bibinfo{journal}{Phys. Rev. Lett.} \textbf{\bibinfo{volume}{83}},
  \bibinfo{pages}{895} (\bibinfo{year}{1999}).

\bibitem[{\citenamefont{Davis and Gardiner}(2002)}]{Davis2001a}
\bibinfo{author}{\bibfnamefont{M.~J.} \bibnamefont{Davis}} \bibnamefont{and}
  \bibinfo{author}{\bibfnamefont{C.~W.} \bibnamefont{Gardiner}},
  \bibinfo{journal}{J. Phys. B} \textbf{\bibinfo{volume}{35}},
  \bibinfo{pages}{733} (\bibinfo{year}{2002}).

\bibitem[{\citenamefont{Gardiner et~al.}(1997)\citenamefont{Gardiner, Zoller,
  Ballagh, and Davis}}]{Gardiner1997b}
\bibinfo{author}{\bibfnamefont{C.~W.} \bibnamefont{Gardiner}},
  \bibinfo{author}{\bibfnamefont{P.}~\bibnamefont{Zoller}},
  \bibinfo{author}{\bibfnamefont{R.~J.} \bibnamefont{Ballagh}},
  \bibnamefont{and} \bibinfo{author}{\bibfnamefont{M.~J.} \bibnamefont{Davis}},
  \bibinfo{journal}{Phys. Rev. Lett.} \textbf{\bibinfo{volume}{79}},
  \bibinfo{pages}{1793} (\bibinfo{year}{1997}).

\bibitem[{\citenamefont{Zhuravlev et~al.}(2000)\citenamefont{Zhuravlev,
  Muryshev, and Fedichev}}]{Zhuravlev2001a}
\bibinfo{author}{\bibfnamefont{O.~N.} \bibnamefont{Zhuravlev}},
  \bibinfo{author}{\bibfnamefont{A.~E.} \bibnamefont{Muryshev}},
  \bibnamefont{and} \bibinfo{author}{\bibfnamefont{P.~O.}
  \bibnamefont{Fedichev}}, \bibinfo{journal}{Phys. Rev. A}
  \textbf{\bibinfo{volume}{64}}, \bibinfo{pages}{053601}
  (\bibinfo{year}{2000}).

\bibitem[{\citenamefont{Gardiner et~al.}(2002)\citenamefont{Gardiner, Anglin,
  and Fudge}}]{Gardiner2002a}
\bibinfo{author}{\bibfnamefont{C.~W.} \bibnamefont{Gardiner}},
  \bibinfo{author}{\bibfnamefont{J.~R.} \bibnamefont{Anglin}},
  \bibnamefont{and} \bibinfo{author}{\bibfnamefont{T.~I.~A.}
  \bibnamefont{Fudge}}, \bibinfo{journal}{J. Phys. B}
  \textbf{\bibinfo{volume}{35}}, \bibinfo{pages}{1555} (\bibinfo{year}{2002}).

\bibitem[{\citenamefont{Williams et~al.}(2002)\citenamefont{Williams, Zaremba,
  Jackson, Nikuni, and Griffin}}]{Williams2002a}
\bibinfo{author}{\bibfnamefont{J.~E.} \bibnamefont{Williams}},
  \bibinfo{author}{\bibfnamefont{E.}~\bibnamefont{Zaremba}},
  \bibinfo{author}{\bibfnamefont{B.}~\bibnamefont{Jackson}},
  \bibinfo{author}{\bibfnamefont{T.}~\bibnamefont{Nikuni}}, \bibnamefont{and}
  \bibinfo{author}{\bibfnamefont{A.}~\bibnamefont{Griffin}},
  \bibinfo{journal}{Phys. Rev. Lett.} \textbf{\bibinfo{volume}{88}},
  \bibinfo{pages}{070401} (\bibinfo{year}{2002}).

\bibitem[{\citenamefont{Feder et~al.}(1999{\natexlab{b}})\citenamefont{Feder,
  Clark, and Schneider}}]{Feder1999a}
\bibinfo{author}{\bibfnamefont{D.~L.} \bibnamefont{Feder}},
  \bibinfo{author}{\bibfnamefont{C.~W.} \bibnamefont{Clark}}, \bibnamefont{and}
  \bibinfo{author}{\bibfnamefont{B.~I.} \bibnamefont{Schneider}},
  \bibinfo{journal}{Phys. Rev. Lett.} \textbf{\bibinfo{volume}{82}},
  \bibinfo{pages}{4956} (\bibinfo{year}{1999}{\natexlab{b}}).

\bibitem[{\citenamefont{Tsubota et~al.}(2002)\citenamefont{Tsubota, Kasamatsu,
  and Ueda}}]{Tsubota2002a}
\bibinfo{author}{\bibfnamefont{M.}~\bibnamefont{Tsubota}},
  \bibinfo{author}{\bibfnamefont{K.}~\bibnamefont{Kasamatsu}},
  \bibnamefont{and} \bibinfo{author}{\bibfnamefont{M.}~\bibnamefont{Ueda}},
  \bibinfo{journal}{Phys. Rev. A} \textbf{\bibinfo{volume}{65}},
  \bibinfo{pages}{023603} (\bibinfo{year}{2002}).

\bibitem[{\citenamefont{Donnelly}(1991)}]{Donnelly1991a}
\bibinfo{author}{\bibfnamefont{R.~J.} \bibnamefont{Donnelly}},
  \emph{\bibinfo{title}{Quantized vortices in helium {II}}}
  (\bibinfo{publisher}{Cambridge~U.~P.}, \bibinfo{address}{Cambridge},
  \bibinfo{year}{1991}).

\bibitem[{Pen()}]{Penckwitt2002a}
\bibinfo{note}{MPEG movies showing these results are available from}
  \urlprefix\url{http://www.physics.otago.ac.nz/bec/theory.htm}.

\end{thebibliography}

\end{document}